\documentclass[prl,floatfix,showpacs,amsmat,twocolumn]{revtex4}
\usepackage{amssymb}
\usepackage{graphicx}
\def\ket #1{\vert #1\rangle}
\def\bra #1{\langle #1\vert}
\def\ketbra #1 #2 {\langle #1\vert #2 \rangle}
\newcommand{\eps}{\varepsilon}
\newcommand{\be}{\begin{eqnarray}}
\newcommand{\ee}{\end{eqnarray}}
\newcommand{\bea}{\begin{eqnarray}}
\newcommand{\eea}{\end{eqnarray}}
\newcommand{\bma}{\begin{subequations}}
\newcommand{\ema}{\end{subequations}}

\begin{document}

\title{ Quantum simulators, continuous-time automata, and translationally
invariant systems.}

\author{K. G. H. Vollbrecht$^1$ and J. I. Cirac$^1$}
\affiliation{$1$ Max-Planck Institut f\"ur Quantenoptik,
Hans-Kopfermann-Str. 1, Garching, D-85748, Germany }

\date{\today}

\begin{abstract}
The general problem of finding the ground state energy of lattice
Hamiltonians is known to be very hard, even for a quantum
computer. We show here that this is the case even for
translationally invariant systems. We also show that a quantum
computer can be built in a 1D chain with a fixed, translationally
invariant Hamitonian consisting of nearest--neighbor interactions
only. The result of the computation is obtained after a prescribed
time with high probability.
\end{abstract}

\maketitle

The difficulty of simulating the dynamics of quantum systems by
classical means was recognized by Feynman \cite{00} more than two
decades ago. He proposed to use another system, a quantum
simulator, to overcome this problem, introducing several visionary
ideas about quantum computation. At the same time, very powerful
classical techniques have been discovered, which allows us to
tackle important problems in many-body physics. One of the most
important questions in this context is to decide which kind of
problems can (or cannot) be efficiently simulated by quantum (or
classical) simulators.

While it is clear that such a device can simulate the dynamics
very efficiently \cite{LloydScience}, it seems that it cannot be
used to prepare ground states of arbitrary nearest neighbor
interacting Hamiltonians. In fact, very recently it has been shown
that if that was possible even in one spatial dimension \cite{11},
then a quantum computer would be able to solve all NP (and even
QMA) problems efficiently, something which seems to be
unreasonable. One may argue that systems in Nature are not so
general since they typically posses certain symmetries (like
homogeneity, or, equivalently, translational invariance) which
restrict very much the Hamiltonians we are interested in and thus
it is not so surprising that some of them cannot be efficiently
simulated.

Another related question is whether a particular quantum simulator
(eg. one that is translationally invariant (TI) and with nearest
neighbor interactions only) may be as powerful as a general one
\cite{ChristinaKraus}. This question can be answered in the
positive if one shows that it can perform every quantum
computation efficiently (with a polynomial overhead in terms of
the number of qubits). In fact, we have previously shown that this
is the case if one is able to change the evolution Hamiltonian
with time \cite{v1, v2}. Very recently, Werner et al. \cite{QCA}
have also shown that it is possible by alternating two kind of
discrete gates, a result which generalizes previous ones on
cellular automata. However, it still remains to be seen if with a
fixed Hamiltonian that is TI and which only includes nearest
neighbor interactions it is still possible to perform arbitrary
quantum computations efficiently.

In this paper we address the two problems mentioned above. First,
we show that by demanding that a 1D Hamiltonian is TI, a quantum
computer is not more efficient in preparing its ground state. Note
that in order to be able to accommodate the number of parameters
which define QMA problems, we must relax the condition of only
nearest--neighbor interactions, although we still keep only
three--body (sites) interactions. Thus, our results imply that the
homogenity that typically appears in Nature is not enough to make
it simulatable. Second, we show that it is possible to build a
quantum computer based on a static, TI Hamiltonian with only
nearest--neighbor interactions, as long as one is able to prepare
arbitrary product states and measure the sites independently. This
result extends previous ones on cellular automata, but as opposed
to those, has no simple classical analogue since we deal here with
continuous time evolution. The TI quantum computation scheme can
be seen as a combination of a cellular automata approach with  a
continuous-time quantum walk \cite{00,QCA}.

{\it Ground state energy in TI systems:} We start out by showing
that finding the ground state energy $E_0$ of any TI Hamiltonian
in a 1D chain is very difficult (as we increase the number of
sites), even for a quantum computer or simulator. In fact, we show
that for each QMA--complete problem one can find a TI Hamiltonian,
$H_n$, such that: $E_0=0$ if the answer to the problem is 'yes';
$E_0>1/(\text{Poly}(n))$, if the answer is 'no'. Here Poly$(n)$
denotes any polynomial in $n$. This implies that any effective
routine for finding the ground state energy would give us a
possibility to efficiently solve all problems in the complexity
class QMA. Our result heavily relies on the recent discovery
\cite{11} that QMA--complete problems can be encoded in the ground
state energy of a Hamiltonian $h_n$ describing a 1D chain of $n$
system with $d=12$ levels each and with nearest--neighbor
interactions, such that its minimum eigenvalue, $\lambda_{\min}$,
is either zero or $\lambda_{\min}>{\text Poly}(n)$. Our strategy
is to build $H_n$ out of $h_n$ by only increasing the dimension
$d$ by a factor of two.

We take at every site one additional qubit and define the
Hamiltonian $H_n^i=\ket{1}\bra{1}_i \otimes h_n^{(i,\dots,i+n)}$.
That is, if the extra qubit at site $i$ is in the state $\ket{1}$,
then the Hamiltonian $h_n$ is applied to the $(i,\dots,i+n)$
particles in the chain in this order, where we identify the sites
separated by $n+1$ sites. Furthermore, we define the Hamiltonian
$H'$ that only acts one the extra qubits as
$$H'=\frac{1}{n(n-1)} \left[1-\sum_k \ket{1_k}\bra{1_k}+\sum_{k''\neq k'}
\ket{1_{k'} 1_{k''}}\bra{1_{k'} 1_{k''}} \right],$$
where $k,k',k''$ run from site $1$ to $n$. Now, we take
$$H_n=H'+\frac{1}{n}\sum_{i=1}^n H_n^i.$$
This Hamiltonian is translationally invariant, and contains up to
three--body interactions only. It is easy to show that it fulfills
the conditions mentioned above, since its eigenvectors have the
form $\ket{k_1,k_2,\dots} \otimes \ket{\phi}$, where
$\ket{k_1,k_2,\dots}$ is a product state  on the extra qubits only
and $\ket{\phi}$ is a state on the rest of the chain. In case
$h_n$ has a zero eigenvalue, with corresponding eigenvector
$\ket{\phi_0}$, we can construct an eigenvector of $H_n$,
$\ket{1,0,0,\dots,0} \otimes \ket{\phi_0}$, also with $E_0=0$.
Otherwise, it is easy to show that the minimal eigenvalue of $H_n$
is lower bounded by $\min\left( \frac{\lambda_{min}}{n} ,\frac{1}
{n^2(n-1)} \right)$, given that if there is no or more than one
extra qubit in $1$, this is penalized in energy by $H'$, whereas
if there is only one, then this is penalized by $H_n^i$.

{\it Continuous--time automata:} Now, we introduce a programmable
quantum computation scheme for an infinite chain of quantum
systems of dimension $d=30$. The program is encoded into the
initial state, while the time evolution is fixed and given by a
universal TI Hamiltonian with nearest--neighbor interaction only.
We will start by introducing a simple standard quantum computing
scheme. Later on, we will show how this quantum computer can be
simulated by a continuous time evolution.


We consider a simple quantum computer with an $n$--qubit 'hard
disk' and a read/write head, that we call the pointer.
This pointer can be moved such that we can address single qubits.
Furthermore, the pointer has an internal quantum state, a qubit.
To perform any quantum computation we write a program consisting
of five different commands: L) The pointer is moved one site to
the left; R) the pointer is moved one site to the right; S) the
qubit at the position of the pointer and the internal state of the
pointer are swaped; G) a G-Gate \cite{QCA} on those two qubits is
applied. The G-gate allows for arbitrary quantum computations if
it can be applied between any two qubits. In this simple model
this can be accomplished by loading qubits with the S-command into
the internal pointer state, which can then be moved to any other
qubit. We now encode this quantum computer into a higher
dimensional chain in the following way: Every site in the chain
has three registers. A 'qubit' register having dimension $2$. It
acts like the normal qubit of the quantum computer. A 'pointer'
register having dimension $3$. Here we encode the pointer, where
$0$ indicates no pointer and $1/2$ the presence of a pointer with
an internal qubit state. Finally, a 'program' register having
dimension 5: One for 'e=empty' and the rest for commands
$\{L,R,S,G\}$. The total dimension of one site defined this way
matches $30$.

We choose $n$ neighboring sites to be the 'quantum computer',
i.e., the qubit-registers of those sites correspond one to one to
the qubits of the quantum-computer we want to simulate (see Fig.\
1). The pointer registers are in state $|0\rangle$ everywhere,
except for one site which contains $\ket{1}$. The program is
written in the program register, an area to the right of the
'quantum computer', where the commands are arranged in the order
they should be executed from left to right. The rest of the
program registers are filled with $\ket{e}$.

\begin{figure}[t]
\centering
\includegraphics[width=85mm]{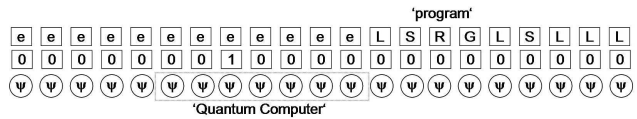}
\label{fig1} \caption{A 7-qubit quantum computer encoded into a
one dimensional chain. The pointer is represented by the $1$, the
program is encoded right of the quantum computer.}
\end{figure}

We assume now a TI Hamiltonian with nearest neighbor interactions,
$H=\sum_i H_i$, where $H_i$ denotes a two-site Hamiltonian acting
only on the sites $i$ and $i+1$ and is given by
\bea H_i=\sum_{C \in \{L,R,S,G\}}\ket{e}_i\bra{C}\otimes
\ket{C}_{i+1}\bra{e} \otimes U^C_{i,i+1}+ h.c. \eea
The first two operators only act on the two program registers at
sites $i$, and $i+1$, respectively, while $U^C$'s are unitary
operators acting  both on the qubit and pointer register at those
two sites. The unitaries $U^S$ and $U^G$ act only on the first
site $i$. A swap resp. G-gate is applied to both the $i$--th qubit
and the internal state of the pointer in case a pointer is
present, and the identity operation otherwise. $U^R$ swaps the two
pointer sites while $U^L$ swaps the two qubit sites. Let us assume
now, that we apply one of those unitaries on every pair of sites
$(i,i+1)$ starting from the right end of the chain going  step by
step to the left. $U^S$ and $U^G$ will do nothing until we reach
the position where a pointer is present in the first of the two
participating sites. In this case, the corresponding gates are
applied. The $U^R$ gate will swap the two pointer states at every
position. As a result the pointer, marked by $\ket{1}$, is moved
one step to the right. Finally, $U^L$ will swap the two qubit
registers. As a result the whole qubit-register chain is moved one
step to the right. Relative to the 'quantum computer' position,
the pointer is moved one step to the left. Therefore,  if such a
sequence of unitaries is applied from right to left the
corresponding set of commands is applied to the 'quantum
computer'. Note, that it is sufficient to start  this sequence of
unitaries at any position to the right of the  'quantum commuter'
and to stop it once they are to its left. Our claim is now that
this is exactly what the time evolution basically does.

To this aim, we consider the time evolution operator
$e^{-iHt}=\sum_n \frac{(-iHt)^n}{n!}$. Applied to the initial
state $\ket{start}$ we just end up in a linear combination of
states of the form $H^n\ket{start}$. Since $H$ is defined as a sum
of terms of the form $h_i^C:=\ket{e_i, C_{i+1}}\bra{C_{i},e_{i+1}}
\otimes U^C_{i,i+1}$ (and the hermitian conjugates), we can write
the result of a time evolution  as a linear combination of states
of the form $h_n\dots h_3h_2h_1\ket{start}$, where every
$h_k\in\{h_i^C, {h_i^C}^\dagger\}$. The effect of any such $h_i^C$
on a state is the following: it either moves the command $C$ from
place $(i+1)$ to $i$ and applies $U^C$ to both sites $(i,i+1)$, or
it maps the state to zero if no $C$ is at ${i+1}$ and no empty
space $\ket{e}$ at $i$. In the same manner ${h_i^C}^\dagger$
results in a state where the command $C$ is moved from $i$ to
$i+1$ while ${U^C}^\dagger$ is applied. Therefore $h_n\dots
h_3h_2h_1\ket{start}$ is either zero, or a state where several of
the commands have moved while applying the assigned unitaries (or
their conjugates) all the way from their initial to their final
position. Note that such a command can only move, if the
corresponding neighboring site is in the state $\ket{e}$. This
implies that, up to some $\ket{e}$'s in between, the order of the
commands in the program register can never change. Further note
that if a command moves to the left and afterwards to the right,
we end up in the same state since $U^C$ and ${U^C}^\dagger$ cancel
each other. This implies that two states $h_n\dots
h_3h_2h_1\ket{start}$, $g_m\dots g_3g_2g_1\ket{start}$ are equal,
iff the configuration of commands in the program registers are
equal.

Assume now that we measure the program registers in the standard
product basis after some time evolution. According to the above
discussion we can conclude just from the configuration of the
program register what has happened to the pointer and qubit
registers. In particular, suppose that we measure a configuration
where all the commands that were initially to the right of the
'quantum computer'(see Fig.\ 1), are found to the left of the
'quantum computer'. In this case the whole program has been
executed (in the right order) to the 'quantum computer' and we can
read out the result. If this is not the case, we can in principle
continue time evolution and repeat until we found a positive
result. To boost the probability of success, we can increase the
program code by some irrelevant code, e.g. by adding $\ket{L}'s$
that do not effect the result of the computation if carried out
after the real program. This irrelevant code will act as a kind of
barrier, that prevents the real program code to move to the right
and forces it to do the computation.

In order to verify that this is an effective way of carrying out
quantum computations, we have to guarantee that the probability of
finding the whole program to the left of the 'quantum computer' is
finite after a finite time (that only grows polynomially in the
number of qubits). To this end we have to solve the above problem
and calculate (a lower bound to) the probability of success. Note,
that given the initial state, the whole system stays in a subspace
that can be labelled by only the configuration of the commands in
the program register. So, we can map our system onto a chain of
qubits, where $\ket{0}$ means 'empty' and $\ket{1}$ means
'command'. Note that, since the order of the commands stays
unchanged under the evolution, we do not have to distinguish
between the different commands. The Hamiltonian  simplifies to
$H=\sum_i H_i$ with $H_i=\ket{1_i 0_{i+1}}\bra{0_i 1_{i+1}}+h.c.$.
At this point we see that the efficiency of the computation does
not depend on the specific program we want to run. Let us take
this system as electrons in a lattice, where  $\ket{1}$ stands for
an 'electron' and $\ket{0}$ for an empty site and $H_i$ is a
hopping term, allowing the electron to hop from one site to a
neighboring site. As usual for fermions, there are no hopping
terms allowed resulting in two electrons sitting at the same site.
Thus, we end up with a system of non--interacting fermions in
second quantization. To solve it, it is simple to go back into
first quantization, where we just have to consider a single
electron Hamiltonian, $H=\sum_k \ket{k+1}\bra{k}+h.c.$, and all
the effects coming from the Pauli principle are automatically
implied by the anti-symmetrization. The single electron problem
can be easily solved. Let us assume $M$ sites with periodic
boundary conditions. The eigenvectors are given by
$\psi_q=\frac{1}{\sqrt{M}}\sum_x e^{i\frac{2pi}{M} xq}\ket{x}$
with corresponding eigenvalues $\eps(q)= 2 \cos(\frac{2\pi}{M}q)$.

Let us now look at $M$ sites where the first $N$ sites are filled
with each one electron. Remember, that these $N$ electrons
correspond to the program and enough irrelevant code to force the
program to go into the desired direction. An electron sitting at
site $y$ is written in terms of the eigenvectors  as
$\phi_y=\frac{1}{M}\sum_q \psi_q
e^{-i\frac{2\pi}{M}yq}=\frac{1}{M}\sum_{q,k}
e^{i\frac{2\pi}{M}(x-y)q} \ket{x}$. After the time evolution, the
state is changed into $\phi_{y,t}=\frac{1}{M}\sum_{q,k}
e^{i\frac{2\pi}{M}(x-y)q+i \eps(q)t} \ket{x}$. We assume now $N$
electrons sitting in the first $N$ sites what leads to the state
$\psi_{0}=S[\ket{\phi_1,\phi_2,\phi_3,\dots,\phi_N}]$, where $S$
denotes the fermionic anti-symmetrization operator. After waiting
a time $t$ we end up in the state $
\psi_{t}=S[\ket{\phi_{1,t},\phi_{2,t},\phi_{3,t},\dots,\phi_{N,t}}]$.
We now want to compute the probability $p_1$ to find particle $1$
at time $t$ still in one of the first $N$ sites. Note that, since
the vectors $\phi_{y,t}$ are still orthogonal, the reduction to
the first particle gives us an averaging over all the starting
positions $y$. This probability yields
\bea \label{p1}
p_1=\frac{1}{N}\sum_{y=1}^{N}\sum_{x=1}^{N}|\ketbra {\phi_{y,t}}
{x} |^2.
\eea
In the Appendix it is shown that by choosing $t$ to be
proportional to $N$, e.g. $t=5000 N$, we can bound this
probability to be smaller than a fixed constant, e.g. $p_1 < 0.3$.
So we can guarantee to find particle $1$ after a polynomial time
with probability bigger than $p=0.7$ outside the starting area.
From this we can calculate the expected number of electrons that
will be found in those areas. They will be $N p_1$ and $N
(1-p_1)=N p$. For a successful computation a fixed number $k$ of
electrons have to leave their starting area. But what is the
probability to find more than $k$ electrons outside the starting
area if we know that the expectation value is $N p$? Let us assume
the worst case scenario: We either get $(k-1)$ electrons with
probability $p_f$ or $N$ electrons with probability $p_s$. Since
we know that $p_f (k-1)+p_s N=N p$ we can conclude that the
success probability $p_s>\frac{1-k+Np}{1-k+N}$ which can be made
arbitrarily close to $p$ by choosing $N$ to grow polynomially with
$k$. Now let us apply this to our model. Note that, due to the
symmetry of the problem, the number of electrons that moved to the
left of the starting position  and to the right of the starting
position will be the same.  We will therefore assume all electrons
moving in the right direction, what can be corrected by an
irrelevant factor of two for all length in the following
discussion. We have a program of length $l_p$ and a quantum
computer of length $l_q$. Instead of of searching for $l_p$
'electrons' to the left of the 'quantum computer' we can search
for $l_p+l_q$ 'electrons' leaving the starting area. The extra
$l_q$ 'electrons' will be just part of the irrelevant code; that
guarantees that the real program completely passed the quantum
computer. Then we choose $N$ to be e.g. $(l_p+l_q)^2$. The above
calculation then tells us that after a time of $5000 (l_p+l_q)^2$
the computation is successful with a probability higher than
$p=0.7$ (if we do not assume the worst case it is quite likely
that we approach 1). Since $l_p$ and $l_q$  grows only polynomial
the same holds for the evolution time.

{\it Conclusion:} We have shown that calculating the  ground state
energy of a translationally invariant Hamiltonians in 1D is as
hard as solving QMA-problems. We have also introduced a
programable quantum computation scheme using one fixed
translationally invariant nearest--neighbor Hamiltonian. The
program is encoded in the initial state. The computation itself
requires only enough patience but no active further control.
 We acknowledges support from EU projects (SCALA and CONQUEST),
DFG-Forschungsgruppe 635, and the excellence clusters MAP and NIM.
K.V. thanks Frank Verstraete for helpful discussions.

{\it Appendix: }
 For equation (\ref{p1}) we get
\bea
p_1=\frac{1}{N M^2} \sum_{x,y=1}^{N}\sum_{q,q'=1}^{M} \cos
\left [{(q-q')(x-y)\frac{2\pi}{M}+ \Delta t} \right]
\eea
where $\Delta=\eps(q)-\eps(q')$. We assume that $N$ and $M$ are
large and $M \gg N$.  If we sum over the cases $q=q'$ we get
$N/M$. Since this term vanishes in the limit of $M \gg N$, we can
ignore those cases. Let us now approximate the sums over $x$ and
$y$ by integrals
\bea \frac{N}{M^2} \sum_{q,q'=1}^{M}
\int_{X,Y=0}^{1} dX dY \cos \left [{(q-q')(X-Y)N \frac{2\pi}{M}+
\Delta t} \right] \eea These integrations leads to \bea \frac{1}{N
\pi^2} \sum_{q,q'=1}^{M} \frac{\cos(\Delta t)
[\sin(\frac{N}{M}(q-q')\pi)]^2}{(q-q')^2}.
\eea
Now let us define $\delta=|q-q'|$. Note that all terms only depend
on the absolute value $q-q'$ and that we already neglected the
terms with $q-q'=0$. We get $ p_1=\frac{1}{N \pi^2}
\sum_{\delta=1}^{M} g(\delta) f(\delta). $ with
$g(\delta)=\frac{[\sin(\frac{N}{M}\delta\pi)]^2}{\delta^2}$ and
$f(\delta)=2\sum_{q=1}^{M-q} cos(\eps(q)-\eps(q+\delta))t$. Note
that for $t=0$ we get that $f(\delta)=2(M-q)$ and in this case the
overall sum converges to $1$. One observation is that the main
contribution to the sum comes from the parts where $\delta$ is
smaller than $2 M/N$. Furthermore for $t>0$  we get $f(\delta)\leq
2(M-q)$. To get a bound we now want to calculate $f(\delta)$ in
the relevant region $\delta <2M/N$. Since $\delta \frac{2 \pi}{M}$
is small in this region, we can approximate
$\eps(q)-\eps(q+\delta)$ by the derivative $\eps'(q) \frac{2
\pi}{M}$ leading to $ f(\delta)=2\sum_{q=1}^{M-q} \cos[
\sin(\frac{2 \pi}{M} q) 2t {\delta 2 \pi/M}]. $ This can be solved
by converting the sum into an integral leading to $f(\delta)=2 M
\text{JBessel}(0, 4t \delta \pi/M))$\cite{mate}. To get an upper
bound we split up the sum in three terms \bea \nonumber
p_1=\sum_{\delta=1}^{\eps M/N} g(\delta)
f(\delta)+\sum_{\delta=\eps M/N}^{2 M/N} g(\delta)
f(\delta)+\sum_{\delta=2 M/N}^{M} g(\delta) f(\delta) \\ \nonumber
\eea Assuming $f(\delta)=2 M$ for the first term we get for small
$\eps$ a bound of $2 \eps$. The second term we get as bound
$\max[f(q)/2M, \eps M/N < q < 2M/N]$. The last term we can
numerical calculate for large $M$ and get it to be smaller then
$0.05$. If we set in $\eps=0.001$, $t=5000N$ we get that $p_1\leq
0.3$.


\end{document}